\documentclass[10pt,journal,compsoc]{IEEEtran}
\ifCLASSOPTIONcompsoc
  \usepackage[nocompress]{cite}
\else
  \usepackage{cite}
\fi
\ifCLASSINFOpdf
  \usepackage[pdftex]{graphicx}
\else
  \usepackage[dvips]{graphicx}
\fi
\usepackage{amssymb}
\usepackage[ruled,vlined,linesnumbered,noresetcount]{algorithm2e}
\usepackage{multirow}

\usepackage{amsmath}
\usepackage{url}
\begin{document}
\title{Learning to Index for Nearest Neighbor Search}

\author{Chih-Yi Chiu,
        Amorntip Prayoonwong,
        and Yin-Chih Liao% <-this % stops a space
\IEEEcompsocitemizethanks{\IEEEcompsocthanksitem C. Y. Chiu, A. Prayoonwong, and Y. C. Liao are with the Department of Computer Science and Information Engineering, National Chiayi University, Taiwan, R.O.C. \protect\\
E-mail: cychiu@mail.ncyu.edu.tw; aprayoonwong@gmail.com; zzzzbenny6209@gmail.com.}% <-this % stops a space
%\thanks{Manuscript received.}
}


\markboth{Journal of \LaTeX\ Class Files}%
{Shell \MakeLowercase{\textit{et al.}}: Bare Advanced Demo of IEEEtran.cls for IEEE Computer Society Journals}



\IEEEtitleabstractindextext{%
\begin{abstract}
In this study, we present a novel ranking model based on learning neighborhood relationships embedded in the index space.
Given a query point, conventional approximate nearest neighbor search calculates the distances to the cluster centroids, before ranking the clusters from near to far based on the distances.
The data indexed in the top-ranked clusters are retrieved and treated as the nearest neighbor candidates for the query.
However, the loss of quantization between the data and cluster centroids will inevitably harm the search accuracy.
To address this problem, the proposed model ranks clusters based on their nearest neighbor probabilities rather than the query-centroid distances.
The nearest neighbor probabilities are estimated by employing neural networks to characterize the neighborhood relationships, i.e., the density function of nearest neighbors with respect to the query.
The proposed probability-based ranking can replace the conventional distance-based ranking for finding candidate clusters, and the predicted probability can be used to determine the data quantity to be retrieved from the candidate cluster.
Our experimental results demonstrated that the proposed ranking model could boost the search performance effectively in billion-scale datasets.
\end{abstract}

% Note that keywords are not normally used for peer review papers.
}

\maketitle

% is not selected <OR> if conference mode is selected - because compsoc
% conference papers position the abstract like regular (non-compsoc)
\IEEEdisplaynontitleabstractindextext

% \fi
\IEEEpeerreviewmaketitle

\ifCLASSOPTIONcompsoc
\IEEEraisesectionheading{\section{Introduction}\label{sec:introduction}}
\else
\section{Introduction}
\label{sec:introduction}
\fi

\IEEEPARstart{N}{earest} neighbor (NN) search in a high-dimensional and large-scale dataset is highly challenging and it has recently attracted much active interest from researchers. This approach is a fundamental technique in many applications, such as classification \cite{zhang2017learning}, feature matching \cite{silpa2008optimised}, recommendation \cite{arampatzis2017suggesting}, and information retrieval \cite{zhang2017fast}.
An exhaustive approach is prohibitive so numerous approximate NN search algorithms have been proposed to address issues in terms of the search accuracy, computation efficiency, and memory consumption.

One solution is to generate compact codes to replace the original data during NN search.
Research into compact code generation has focused on two main areas: binary embedding and codebook learning.
In binary embedding, each data point is transformed into a binary pattern with a set of hashing functions.
In codebook learning, a set of centroids is generated by a clustering technique and each data point is assigned to the cluster with the nearest centroid.
By compiling these codes as a codebook, we can build an index structure to accelerate NN search.
Common indexing approaches include cluster ranking and pruning \cite{manning2008introduction}\cite{huang2010mining}, where we expect a set of similar data is grouped in the same cluster.
At the query time, we rank the clusters based on a similarity or distance metric with respect to the query.
Only data points indexed in high-ranking clusters are retrieved for verification, and those in low-ranking clusters that are considered irrelevant to the query can be filtered out.

\begin{figure}[t]
\begin{center}
\includegraphics[width=0.4\textwidth]{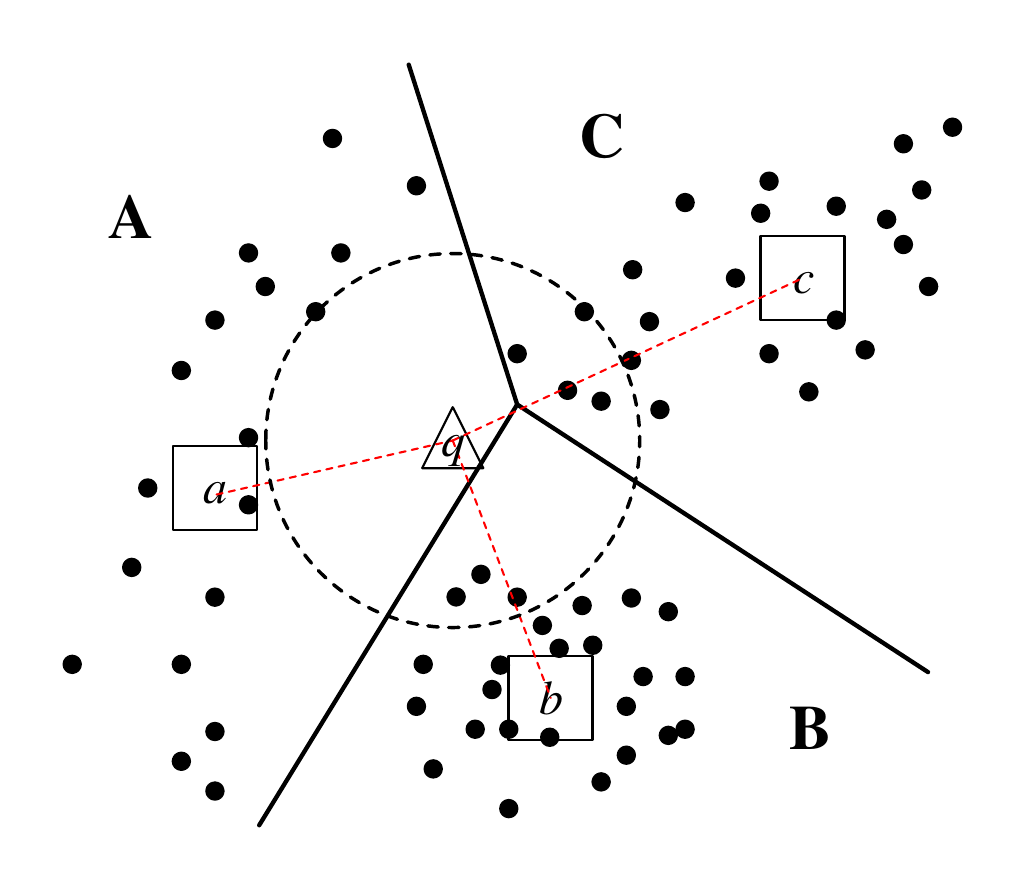}
\caption{A 2D Euclidean space, where \textbf{A}, \textbf{B}, and \textbf{C} are clusters, \textit{a}, \textit{b}, and \textit{c} are the respective centroids, and \textit{q} is the query. The clusters can be ranked based on their query-centroid distances, which is widely used in existing NN search methods. However, the distance-based ranking does not reflect the NN distribution among the clusters well.}
\label{fig_intro}
\end{center}
\end{figure}

Inevitably, the information loss between the original data points and the corresponding quantized codes will impair the indexing performance, although we can increase the number of codewords or use a better quantization method to alleviate it to some extent.
Figure \ref{fig_intro} shows an example of a two-dimensional Euclidean space with three clusters: \textbf{A}, \textbf{B}, and \textbf{C}. Let \textit{a}, \textit{b}, and \textit{c} be the centroids of the respective clusters, and \textit{q} is the query point.
The Euclidean distances between the query and centroids are denoted as $\| \overline{qa} \|$, $\| \overline{qb} \|$, and $\| \overline{qc} \|$.
Assuming that $\| \overline{qa} \| < \| \overline{qb} \| < \| \overline{qc} \|$, then the order of traversing the clusters is from \textbf{A} to \textbf{B} to \textbf{C}.
It should be noted that the traversal order does not reflect the NN distribution among the clusters well.
In fact, cluster \textbf{C} contains more relevant data that are closer to the query (in a certain radius) than the other clusters, and thus it should be visited earlier so more true positives can be retrieved.
However, the existing NN search methods (e.g., cluster ranking and pruning) rely mainly on the Euclidean distances to rank clusters.
Cluster \textbf{C} might not be visited if we rely on the distance-based ranking to traverse only the top one or two clusters, and thus the relevant data indexed in cluster \textbf{C} cannot be found.

In this study, we present a novel ranking model that learns neighborhood relationships embedded in the index space, which can be employed to estimate the NN distribution probabilities among clusters for a given query.
The underlying concept for the proposed model is that the cluster ranking may be determined based on the NN probabilities of clusters rather than their centroid distances, with respect to the query.
As shown in Figure \ref{fig_intro}, cluster \textbf{C} contains most of the NNs to query \textit{q}.
Thus, it can be regarded as the most relevant cluster, so its priority should be at the topmost rank.
In order to characterize the neighborhood relationships, we propose learning from a training set of query-dependent features.
That is, we formulate a nonlinear function of the query-dependent features for predicting the NN probabilities of clusters.
The clusters are ranked according to the NN probabilities instead of the Euclidean distances, so retrieving the high-ranking clusters is expected to obtain more true positives for the query.
Moreover, the predicted probability can be used to calculate the data quantity to be retrieved from the cluster.
Intuitively, we should retrieve more data from the high-ranking clusters to identify true positives, and retrieve less data from the low-ranking clusters to reduce false positives.
Rather than retrieving all of the data indexed in a cluster, the data quantity that needs to be extracted from the cluster is proportional to its NN probability for a better retrieval quality.
The proposed ranking model can be applied hierarchically to learn the neighborhood relationships embedded in a multi-level index space.
It is time- and space-efficient while effectively boosting the search accuracy.

\begin{figure}[t]
\begin{center}
\includegraphics[width=0.5\textwidth]{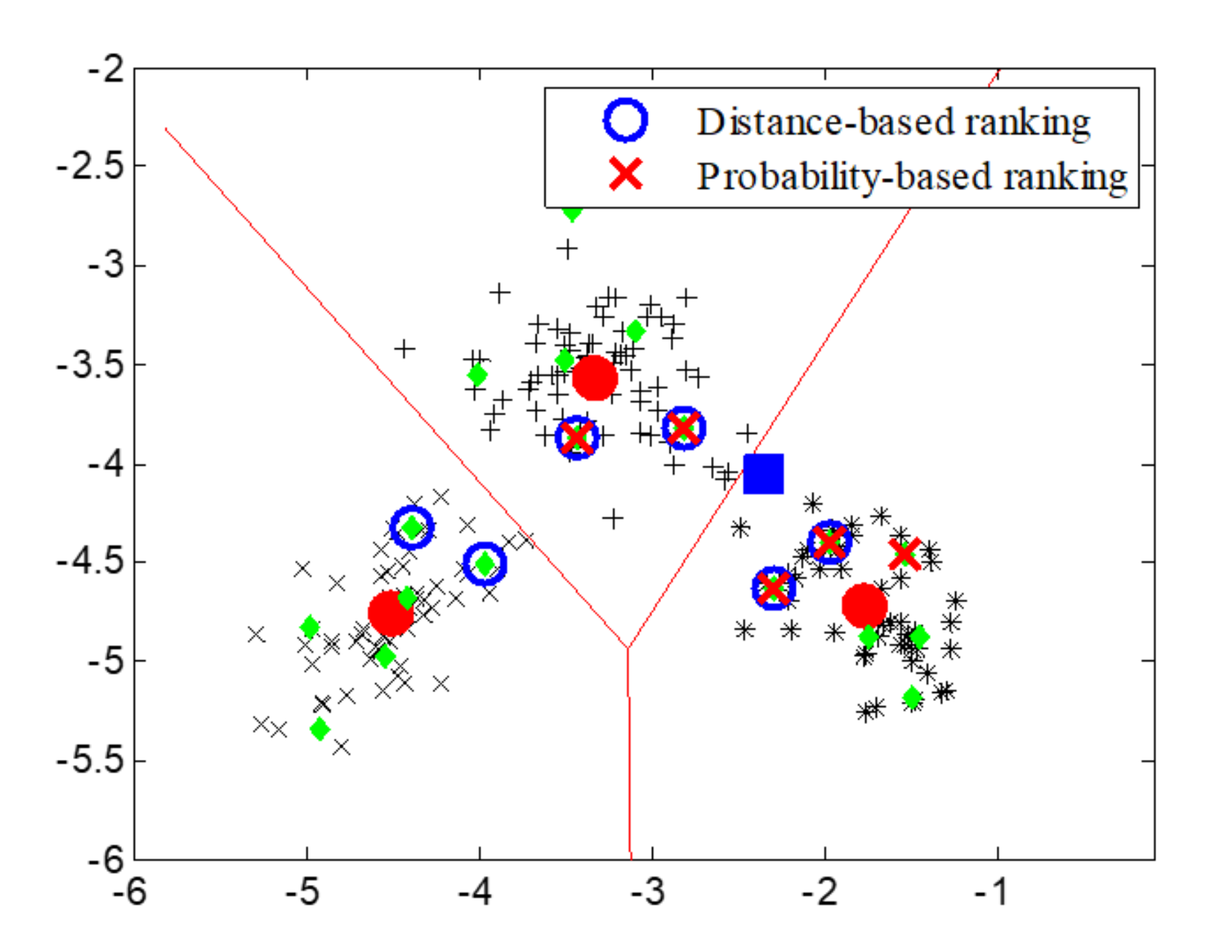}
\caption{Probability-based ranking vs. distance-based ranking. Data points are from the wine dataset (UCI machine learning repository), a thirteen-feature set labeled by three different classes. We project the data points to a 2D space constituted by the top two eigenvectors with the highest eigenvalues for visualization purposes.}
\label{fig_compare}
\end{center}
\end{figure}

Figure \ref{fig_compare} shows an example of the retrieval results obtained by the proposed probability-based ranking and typical distance-based ranking, where the blue rectangle is a query point, large red circles are cluster centroids, and small green diamonds are subcluster centroids.
The results are produced by selecting the subclusters with the highest probabilities/smallest distances to the query.
The probability-based ranking presents more relevant retrieval compared to the distance-based ranking.

The proposed probability-based ranking has the following advantages compared with the distance-based ranking:

\begin{itemize}
\item The probability-based ranking can be used to replace the distance-based ranking during NN search.
We alleviate the information loss problem caused by embedding and quantization by predicting the NN probabilities through nonlinear mappings of query-dependent features to rank clusters and estimate the quantity to be retrieved from each cluster.
The retrieval quality is improved significantly through computation- and memory-efficient indexing.
\item The probability-based ranking can be integrated easily with existing quantization/search methods.
It returns highly relevant data as candidates, which are then reranked through asymmetric distance computation (ADC).
We demonstrated the feasibility and superior performance of the proposed method based on a multi-level index structure comprising two popular quantization methods residual vector quantization (RVQ) and optimized product quantization (OPQ).
\end{itemize}

We conducted evaluation experiments using the billion-scale SIFT and DEEP datasets.
We implemented the cluster ranking function of three query-dependent features and three quantity estimation functions in the probability-based ranking model to compare them with the distance-based ranking model.
The results indicated that the neighborhood relationships can be learned and utilized to improve the ranking quality during NN search.

The remainder of this paper is organized as follows.
In Section 2, we discuss previous research related to NN search techniques.
In Section 3, we explain the proposed probability-based ranking model.
In Section 4, we present the experimental results obtained using large-scale datasets.
Finally, we give our conclusions in Section 5.

\section{Related Work}

NN search has been a highly active research area in recent decades.
In the following, we review some of the fundamental techniques related to our study, including data quantization and data indexing.

\subsection{Data Quantization}

Data quantization techniques generate compact codes transformed from the original data and they can be divided into two categories: binary embedding and codebook learning.
Binary embedding provides the most compact data representation.
In general, the original data are transformed into an intermediate space by a set of hash functions.
The transformed data are then quantized to binary embedding codes.
The Hamming distance between two binary codes can be calculated efficiently with hardware-supported machine instructions.
Numerous learning to hash methods have been proposed recently for learning the data-dependent and task-specific hash functions to generate a short but effective binary code, including unsupervised \cite{gong2013iterative}\cite{ercoli2017compact}\cite{shen2018unsupervised}\cite{hu2018hashing} and supervised \cite{ge2014graph}\cite{li2017learning}\cite{shen2017asymmetric}\cite{yang2018supervised}, to name a few.
A comprehensive study for learning to hash can be found in some survey papers \cite{wang2014hashing}\cite{wang2016learning}\cite{wang2018survey}.
Applying binary embedding can save time and memory but its search accuracy is severely degraded.
Due to the huge quantization loss, the measurement between two binary embedding codes is imprecise with only a few distinct Hamming distances.

The codebook learning approach produces a set of representatives for vector quantization.
One of the most popular algorithms for this purpose is \textit{k}-means clustering, which iteratively partitions the vector space into \textit{k} Voronoi cells according to the data distribution.
Product quantization (PQ) is a generalization of \textit{k}-means \cite{jegou2011product}.
PQ partitions the feature vector of a data point into several segments of disjoint subvectors.
Each segment is encoded by a corresponding codebook, and each data point is represented by a concatenation of segment codewords.
PQ produces huge quantization levels using small-size codebooks, so codebook learning and data quantization can be implemented efficiently.
This becomes a common technique for processing data in the high-dimensional vector space.
For example, PQk-means \cite{matsui2017pqk} employs PQ codes to enable fast and memory-efficient \textit{k}-means clustering.
Blalock and Guttag \cite{blalock2017bolt} presented a variation of PQ with fast encoding speed based on small codebooks, and efficient distance computation by learning a quantization function that approximates the query-centroid distances in lookup tables.
Xu et al. \cite{xu2018online} proposed an online PQ method that can update the quantization codebook dynamically based on the incoming streaming data and by using insertion/deletion operations to reflect the real-time data behavior.

To reduce the quantization distortions in PQ, some studies searched for the optimal space decomposition to learn codebooks.
In OPQ \cite{ge2013optimized} and Cartesian \textit{k}-means \cite{norouzi2013cartesian}, generalized PQ is applied by finding the optimal transformation matrix and sub-codebooks in the data space to minimize the quantization distortion from a global viewpoint.
Locally OPQ (LOPQ) \cite{kalantidis2014locally} assumes the data distribution in each cluster is unimodal, so an individual OPQ is performed per cluster with a lower distortion.
Bilinear optimized PQ \cite{yu2017bilinear} learns a bilinear projection for PQ, which exploits the natural data structure to reduce the time and space complexity.

The PQ-based methods try to decompose the data space into orthogonal subspaces, but an alternative approach called non-orthogonal quantization approximated the data based on the sum of the codewords instead of concatenation.
For example, the widely used RVQ method \cite{chen2010approximate}\cite{wei2014projected}\cite{babenko2016efficient} quantizes the displacement between the original vector and the nearest centroid, i.e., the residual vector.
Encoding the residual usually yields a lower quantization error than encoding the original vector.
Additive quantization \cite{babenko2014additive} eliminates the orthogonal subspace assumption and represents a vector with a sum of codewords to achieve considerably lower approximation error.
Similarly, composite quantization \cite{zhang2014composite} approximates a vector using the summation of codewords under the constraint of a constant inter-dictionary-element-product.
Tree quantization \cite{babenko2015tree} uses a tree graph of codebooks, which is constructed by minimizing the compression error through integer programming-based optimization.
AnnArbor \cite{babenko2017annarbor} employs arborescence graphs to encode vectors based on their displacements between nearby vectors.

\subsection{Data Indexing}

The tree-based index structure is one of the most widely used techniques for vector space indexing \cite{beygelzimer2006cover}\cite{nister2006scalable}\cite{silpa2008optimised}.
However, it usually requires a considerable amount of memory to store a huge index structure and the search time grows rapidly with the dimensionality.
To address this problem, Muja and Lowe \cite{muja2014scalable} presented the priority search \textit{k}-means tree algorithm for effective matching in high-dimensional spaces.
Houle and Nett \cite{houle2015rank} proposed the rank cover tree, which uses the ordinal ranks of the query distances to prune data points.
Liu et al. \cite{liu2017generalized} introduced an aggregating tree, which is a radix tree built based on RVQ encodings, and the beam search algorithm is applied to search for NNs.

Inverted file system with ADC (IVFADC) is another popular framework, which was designed to handle billion-scale datasets efficiently \cite{jegou2011product}.
This system employs \textit{k}-means clustering for coarser quantization and then utilizes PQ for residual quantization.
ADC \cite{dong2008asymmetric}\cite{andre2017accelerated}\cite{gordo2014asymmetric} is used for fast Euclidean distance computation via inverted table lookup.
In ADC, the query is kept in its original space and the reference data are kept in the quantized space.
There is no quantization loss at the query side, so ADC can yield a more accurate evaluation compared with the computation between the quantized query and reference data points.
LOPQ \cite{kalantidis2014locally} can be regarded as a modified form of IVFADC that uses separate local PQ codebooks for data compression.
Local codebooks can model the local data distribution more precisely, so its accuracy is better than that of IVFADC.
Baranchunk et al. \cite{baranchuk2018revisiting} proposed grouping and pruning procedures to improve IVFADC, where each cluster is split into subclusters, which form a set of convex combinations of the local neighboring centroids.

The use of multiple quantizers, known as multiple hash tables, can achieve good recall and speed in practice.
For example, \textit{k}-means locality sensitive hashing \cite{pauleve2010locality} uses several codebooks, which are generated by running \textit{k}-means clustering several times for multiple hashing.
The joint inverted indexing method \cite{xia2013joint} creates codewords jointly by single \textit{k}-means clustering, before assigning the codewords to different codebooks so the total distortion of the quantizers is minimal.
Some PQ-based indexing methods exploit the power of PQ in multiple hashing.
The inverted multi-index (IMI) method \cite{babenko2012inverted} constructs a multi-dimensional index table, where the cell centroids are made from the Cartesian product of the codebooks.
A large number of cells can provide very dense partitioning of the space, so IMI can achieve a high recall by traversing only a small fraction of the dataset.
PQTable \cite{matsui2015pqtable} was proposed to accelerate exhaustive PQ search by using the concatenation of PQ codes from multiple hash tables to index a data point.
Multi-index voting \cite{chiu2017approximate} employs a voting mechanism to generate high-quality NN candidates by traversing across multiple hash tables derived from the intermediate spaces.

\section{Probability-Based Ranking}

Given a reference dataset of \textit{I} \textit{l}-dimensional real-valued vectors $V = \lbrace v_i \in \lbrace \mathbb{R} \rbrace ^l | i = 1, 2, ..., I \rbrace$, we construct an index structure with \textit{M} clusters $\left\{ C_m | m = 1, 2, ..., M \right\}$ and a codebook of centroids $\left\{ u_m | m = 1, 2, ..., M \right\}$.
The \textit{m}th cluster $C_m$ is associated with the centroid $u_m$ and an inverted list that contains the reference data indexed in the \textit{m}th cluster expressed as:
  \begin{equation}
  C_m = \lbrace v_i | Z(v_i) = m \rbrace,
  \end{equation}
where function $Z(v_i)$ returns the nearest centroid id of $v_i$:
  \begin{equation}
  Z(v_i) = \underset{m \in \{1, 2, ..., M\}}{\arg\min} \left( \left\| v_i - u_m \right\|, 1 \right),
  \end{equation}
where $\left\| \cdot \right\|$ denotes the 2-norm (Euclidean distance) and $\underset{b}{\arg\min}(B, k)$ returns the \textit{k} arguments indexed by \textit{b} for the \textit{k} minimums in set \textit{B}.
Given a query \textit{q}, a typical approach for utilizing the index structure involves retrieving the top-\textit{R} ranked clusters whose centroids are nearest to \textit{q}.
Denote the top-\textit{R} clusters as $\left \{ C_{m_1}, C_{m_2}, ..., C_{m_R} \right \}$ and the index subscripts are expressed by:
  \begin{equation} \label{eq:dist_ranking}
  \left\{ m_1, m_2, ..., m_R \right\} = \underset{m \in \{1,2,...,M\}}{\arg\min} \left( \left\| q - u_m \right\|, R \right),
  \end{equation}
where $C_{m_r}$ is the \textit{r}th-nearest cluster to \textit{q}.
These clusters are traversed sequentially to retrieve NN candidates from their inverted lists.
This indexing process is known as cluster ranking and pruning \cite{manning2008introduction} and it is usually applied as a coarse filter, which is employed widely in existing NN search methods.
However, as mentioned earlier, the quantization loss in the index space will degrade the retrieval quality.
In order to address this issue, we learn how to model the neighborhood relationships embedded in the index space.

\subsection{Learning to Index} \label{sec:learning_to_index}

We model the neighborhood relationships by a function \textit{f}: $ f(X) = \left\{ p_1, p_2, ..., p_M \right\}$ , which maps a query-dependent feature vector \textit{X} to the NN probabilities $\left\{ p_1, p_2, ..., p_M \right\}$, where $p_m$ represents the NN probability of the \textit{m}th cluster.
We then rank the clusters based on their NN probabilities instead of the Euclidean distances in order to output the top-\textit{R} clusters.
Thus, Eq. (\ref{eq:dist_ranking}) is rewritten as:
  \begin{equation} \label{eq:prob_ranking}
  \left\{ m_1, m_2, ..., m_R \right\} = \underset{m \in \{1,2,...,M\}}{\arg \min} \left( p_m, R \right).
  \end{equation}
The cluster sequence ranked by the NN probabilities can reflect the neighborhood relationships between the query and clusters better than that ranked based on the Euclidean distances, thereby improving the retrieval quality.

The mapping function $f(X)$ is constructed in the following training process.
Let $Q=\{ q^{(1)}, q^{(2)}, ... , q^{(T)} \}$ be the training dataset of \textit{T} queries.
The \textit{t}th query $q^{(t)}$ is associated with the weighted ground truth of \textit{K} NNs, denoted as $G^{(t)} = \left\{ g_1^{(t)}, g_2^{(t)}, ..., g_K^{(t)} \right\}$ and $W^{(t)} = \left\{ w_1^{(t)}, w_2^{(t)}, ..., w_K^{(t)} \right\}$, where $g_k^{(t)} \in V$ is the \textit{k}th NN of $q^{(t)}$ and the corresponding weight is $w_k^{(t)}$.
For input \textit{X}, we propose three query-dependent features, comprising the raw query data, query-centroid similarities, and their concatenation.
The raw query data feature is $q^{(t)}$ itself.
The query-centroid similarity feature is derived from a distance feature $D^{(t)} = \left\{ d_1^{(t)}, d_2^{(t)}, ..., d_M^{(t)} \right\}, d_m^{(t)} = \| q^{(t)} - u_m \|$.
Then, the distance feature is transformed into a similarity feature $A^{(t)} = \left\{ a_1^{(t)}, a_2^{(t)}, ..., a_M^{(t)} \right\}$, where $a_m^{(t)}$ is defined by:
  \begin{equation}
  a_m^{(t)} = \frac{\max \left( D^{(t)} \right) - d_m^{(t)}}{\max \left( D^{(t)} \right)}.
  \end{equation}
The mapping function output, also known as the target, is a vector of NN probabilities of \textit{M} clusters, denoted as $Y^{(t)} = \left \{ y_1^{(t)}, y_2^{(t)}, ..., y_M^{(t)} \right \}$.
$y_m^{(t)}$ is defined by:
  \begin{equation} \label{eq:target_y_m}
  y_m^{(t)} = \frac{\sum\limits_{k} \left\{ w_k^{(t)} \big | g_k^{(t)} \in C_m \right\}}{\sum\limits_{k} \left\{ w_k^{(t)} \big | g_k^{(t)} \in C_m \right\} + |C_m|},
  \end{equation}
where $| \cdot |$ denotes the set cardinality.
Then $y_m^{(t)}$ is normalized by sum to one.
Consequently, the \textit{t}th training data $\{ q^{(t)}, G^{(t)}, W^{(t)} \}$ is converted to the input-output pair $\{ X^{(t)}, Y^{(t)} \}$ of \textit{f}.

We employ a fully-connected neural network to learn the neighborhood relationships based on the training dataset $\{ X^{(t)}, Y^{(t)} | t = 1, 2, ..., T \}$.
The input layer receives $X^{(t)}$ and the output layer predicts NN probabilities for \textit{M} clusters, denoted as $P^{(t)} = \left( p_1^{(t)}, p_2^{(t)}, ..., p_M^{(t)} \right)$.
Based on the loss between the predictions $P^{(t)}$ and the target $Y^{(t)}$, we compute the error derivative with respect to the output of each unit, which is backward propagated to each layer in order to tune the weights of the neural network.

\subsection{Quantity Estimation}

By using the learned mapping function, we can predict the NN probability distribution among clusters for the given query.
The NN probabilities can be utilized to rank clusters but also to estimate the data quantities that needs to be retrieved from the clusters.
If we control the amount of data that needs to be retrieved from the clusters, it is reasonable to retrieve more data from the high-probability clusters and less data from the low-probability clusters for the sake of better precision.
To illustrate this idea, we formulate the quantity estimation problem in a two-level index structure.
The first level comprises \textit{M} clusters $\left\{ C_m | m = 1, 2, ..., M \right\}$.
Each $C_m$ is further partitioned by \textit{N} subclusters, i.e., the second-level clusters, denoted as $\left\{ C_m^n | n = 1, 2, ..., N \right\}$.
Thus, the data space is divided into $M \times N$ partitions.
Suppose that we obtain the NN probabilities of the first-level clusters $\left\{ p_m | m = 1, 2, ..., M \right\}$.
Let $\gamma(p)$ be a monotonically increasing function and \textit{T} is the total amount of the second-level clusters that need to be retrieved.
The number of the second-level clusters retrieved from $C_m$ is calculated by:
  \begin{equation} \label{eq:num_sec_cluster}
  S = \max \left( \left\{ N,\mathrm{round} \left( \frac{\gamma (p_m)}{\sum_{1}^{m}\gamma (p_m)} \cdot T \right) \right\} \right),
  \end{equation}
where function $\mathrm{round}(\cdot)$ returns the closest integer. 
Thus, from $C_m$, we select \textit{S} second-level clusters (at maximum of \textit{N}) that are closest to query \textit{q}.
The selection of the top-\textit{S} second-level clusters is dependent on the underlying index structure and we explain it in next subsection.

$\gamma(p)$ can be considered a normalization function that rescales \textit{p} based on some assumed distribution for the NN probabilities among the top-\textit{R} first-level clusters $\left\{ C_{m_1}, C_{m_2}, ..., C_{m_R} \right\}$.
Let $p_{m_r}$ be the NN probability of the \textit{r}th ranked cluster $C_{m_r}$.
The possible normalization functions considered in this study are as follows.
\begin{itemize}
\item \emph{Sum.} Normalized by sum to one:
  \begin{equation} \label{eq:sum_norm}
  \gamma_{sum} \left ( p_{m_r} \right ) = \frac{p_{m_r}}{\sum_{j=1}^{R}p_{m_j}}.
  \end{equation}
\item \emph{Standard.} Rescaled under a normal distribution with a mean of 0 and a standard deviation of 1:
  \begin{equation} \label{eq:std_norm}
  \gamma_{std} \left ( p_{m_r} \right ) = \frac{p_{m_r}-\mu}{\sigma},
  \end{equation}
where $\mu$ and $\sigma$ defined as the mean and standard deviation of the NN probabilities of the top-\textit{R} first-level clusters, respectively.
\item \emph{Exponent.} Defined as a probability density function of the exponential distribution:
  \begin{equation} \label{eq:exp_norm}
  \gamma_{exp} \left ( p_{m_r} \right ) = e^{-\frac{1-p_{m_r}}{\rho}},
  \end{equation}
where $\rho$ is the inverse to the mean of the probabilities of the top-\textit{R} first-level clusters:
  \begin{equation}
  \rho = 1 - \frac{1}{R}\sum_{j=1}^{R}p_{m_j}.
  \end{equation}
\end{itemize}
Our evaluations for these normalization functions are presented in the experimental section.

\subsection{Indexing and Search Algorithm}

In this subsection, we describe the general indexing and search algorithm integrated with the proposed ranking model, and we illustrate the capability based on some hash-based index structures for NN search.
The steps are summarized in Algorithm \ref{alg:index_search}.
In step 1, for the given query \textit{q}, we generate the query-dependent feature \textit{X}.
The trained mapping function $f(X)$ is used to predict the NN probabilities of the first-level clusters $\left\{ p_1, p_2, ..., p_M \right\}$.
The top-\textit{R} first-level clusters $\left\{ C_{m_1}, C_{m_2}, ..., C_{m_R} \right\}$ with the highest NN probabilities are selected.
In steps 4 and 5, we select the top-\textit{S} second-level clusters $\left\{ C_{m_r}^{n_1}, C_{m_r}^{n_2}, ..., C_{m_r}^{n_S} \right\}$ from each $C_{m_r}$.
The reference data indexed in the selected second-level clusters are retrieved as candidates.
We then calculate the distance between the query and each of the candidates for reranking, where ADC is employed and sort the candidates in ascending order and return \textit{k} NNs in the last step.

\begin{algorithm}[t]
  \caption{Indexing and Search}
  \label{alg:index_search}
  \KwIn{first-level clusters $\{ C_m \}$; second-level clusters $\{ C_m^n \}$; mapping function \textit{f}; query \textit{q};}
  \KwOut{\textit{k} NNs of \textit{q};}
  Generate the query-dependent feature vector \textit{X} for \textit{q}\;
  Predict the NN probabilities of the \textit{M} first-level clusters $f(X) = \left\{ p_1, p_2, ..., p_M \right\}$\;
  Select the top-\textit{R} first-level clusters $\left\{ C_{m_1}, C_{m_2}, ..., C_{m_R} \right\}$ by Eq. (\ref{eq:prob_ranking})\;
  For each $C_{m_r}, r \in [1, R]$, calculate the number of the second-level clusters \textit{S} to be retrieved using Eq. (\ref{eq:num_sec_cluster})\;
  For each $C_{m_r}, r \in [1, R]$, select the top-\textit{S} second-level clusters $\left\{ C_{m_r}^{n_1}, C_{m_r}^{n_2}, ..., C_{m_r}^{n_S} \right\}$ and generate a candidate set \textbf{A}\;
  Perform ADC between \textit{q} and each candidate in \textbf{A}\;
  Sort the asymmetric distances in ascending order and return the \textit{k} NNs\;
\end{algorithm}

%\begin{table}[ht]
%\centering
%\caption{Time complexity of Algorithm \ref{alg:index_search}}
%\label{tb:time_complexity}
%\begin{tabular}{ |c|c| }
%  \hline
%  Step 1  &  $O(lM)$ \\
%  \hline
%  Step 2  &  $O(M^2)$ \\
%  \hline
%  Step 3  &  $O(M\lg R)$ \\
%  \hline
%  Step 4  &  $O(R)$ \\
%  \hline
%  Step 5  &  $O(RN\lg S)$ \\
%  \hline
%  Step 6  &  $O(l \beta + \alpha | \mathbf{A} |)$ \\
%  \hline
%  Step 7  &  $O(| \mathbf{A} | \lg k)$ \\
%  \hline
%\end{tabular}
%\end{table}

The time complexity of Algorithm 1 is mainly from the three parts: NN probability prediction (step 2), candidate selection (steps 3-5), and reranking (steps 6-7).
The NN probabilities of the first-level clusters are predicted by the fully-connected neural network.
Suppose the network has $\tau$ hidden layers, each of which contains $\lambda$ units.
The prediction thus spends $O(\tau\lambda^2 + \lambda M)$ time.
Candidate selection requires to sort \textit{M} (\textit{N}) clusters.
We apply partial sorting (e.g., std::nth\_element from C++ STL) to find top clusters with linear complexity, taking $O(M)$ $\big( O(N) \big)$ time in average.
The main concern of reranking is the number of candidates retrieved.
For each candidate, we perform ADC to the query by table lookup.
If we suppose that there are $\alpha$ tables and each table contains $\beta$ codewords representing an $\frac{l}{\alpha}$-dimensional vector, then ADC requires $O(l \beta + \alpha | \mathbf{A} |)$ time.

The memory required by Algorithm \ref{alg:index_search} is mainly for the reference dataset and index structure, which must be loaded into memory for real-time responses in search applications.
The reference data are stored in the form of PQ codes.
Each data point takes $\alpha \lg \beta$ bits in memory and $\frac{\alpha \lg \beta}{8} \cdot I$ bytes for a total of \textit{I} reference data points, with an additional PQ codebook of $8 \beta l$ bytes.
The index structure size is determined by the number of the indexed data points, index table, prediction network, and index codebook.
Assume the number of reference data points \textit{I} is no more than $2^{32}$, we can use 4 bytes to encode the data identity, and $4I$ bytes are required for \textit{I} reference data points.
A two-level index table associates data identities through 64-bit pointers for inverted indexing, which take $8MN$ bytes.
The prediction network occupies $4(\tau\lambda^2 + M\lambda)$ bytes, where the network coefficients are represented by 4-byte floating-point numbers.
The index codebook size is dependent on the underlying quantization method, as discussed in Section \ref{sec:codebook_size}.

The index structure and quantization method determine the cluster selection and candidate generation process specified in step 5 of Algorithm \ref{alg:index_search}.
In next subsections, we take two quantization methods, comprising RVQ and OPQ, to show how to integrate with the proposed ranking model.

\subsubsection{RVQ}

We use the multi-level RVQ method \cite{chen2010approximate}\cite{wei2014projected}\cite{babenko2016efficient}, where multiple quantizers are concatenated sequentially to approximate the quantization errors in the preceding levels.
Each quantizer has its corresponding codebook.
We recall that the first-level cluster set $\left\{ C_m | m = 1, 2, ..., M \right\}$ is associated with the quantizer $Z(v_i)$ and codebook $\left\{ u_m | m = 1, 2, ..., M \right\}$.
The residual of the reference data point $v_i$ known as the quantization error is expressed as:
  \begin{equation}
  e_i = v_i - Z(v_i).
  \end{equation}
Given the residual set $\left\{ e_i | i = 1, 2, ..., I \right\}$, we apply \textit{k}-means clustering to divide the residual space into \textit{N} partitions to generate the residual codebook $\left\{ u^n | n = 1, 2, ..., N \right\}$.
The residual codebook is shared by all of the vectors in the residual space.
In total, \textit{MN} distinct quantization codes are provided by the two codebooks.
The second-level cluster is denoted as $C_m^n$, $m \in [1, M]$ and $n \in [1, N]$:
  \begin{equation}
  C_m^n = \left\{ v_i | Z(v_i) = m, E(e_i) = n \right\},
  \end{equation}
where function $E(e_i)$ returns the nearest centroid id of the residual codebook to $e_i$.
$v_i$ is indexed in $C_m^n$ and it can be approximated by the second-level centroid $u_m^n$:
  \begin{equation}
  v_i \approx u_m^n = u_m + u^n, \forall v_i \in C_m^n.
  \end{equation}
If we want to create a multi-level index structure \cite{wei2014projected}, the codebooks can be created in the same manner.

Suppose that we obtain the top-\textit{R} first-level clusters $\left\{ C_{m_1}, C_{m_2}, ..., C_{m_R} \right\}$ for query \textit{q}.
We want to find the \textit{s}th closest-distance second-level cluster from the \textit{r}th highest-probability first-level cluster, denoted as $C_{m_r}^{n_s}, r \in [1, R], s \in [1, S]$ (step 5 of Algorithm \ref{alg:index_search}).
From each $C_{m_r}$, we select the top-\textit{S} second-level clusters $\left\{ C_{m_r}^{n_1}, C_{m_r}^{n_2}, ..., C_{m_r}^{n_S} \right\}$ (recall that \textit{S} is determined by Eq. (\ref{eq:num_sec_cluster})), where the index superscripts $\left\{ n_1, n_2, ..., n_S \right\}$ are expressed by:
  \begin{equation} \label{eq:residual_ranking}
  \left\{ n_1, n_2, ..., n_S \right\} = \underset{n \in \{1, 2, ..., N\}}{\arg\min} \left( \left\| q - u_{m_r}^n \right\|, S \right).
  \end{equation}
The data indexed in $\left\{ C_{m_r}^{n_s} | r = 1, 2, ..., R, s = 1, 2, ..., S \right\}$ are considered to be the candidates.

\subsubsection{OPQ}

The objective of OPQ is to find the optimal space decomposition for minimum quantization distortions \cite{ge2013optimized}.
The optimal space decomposition is jointly determined by an orthonormal matrix and sub-codebooks, which can be solved by either non-parametric or parametric (based on a Gaussian distribution) optimization.
Following the same process employed for RVQ, we generate the first-level cluster set $\left\{ C_m \right\}$ and create the residual space with the residual data $\left\{ e_i \right\}$.
The residual space is transformed by an orthonormal matrix $\Phi$ and divided into $\alpha$ subspaces.
Each subspace has a sub-codebook containing $\beta$ sub-codewords.
This process is equivalent to generating a residual codebook $\left\{ u^n | n = 1, 2, ..., N \right\}$ in the residual space, where $N = \beta ^ \alpha$, and $u^n$ is represented as the concatenation of sub-codewords from the $\alpha$ subspaces.
The second-level cluster $C_m^n$ is expressed as:
  \begin{equation}
  C_m^n = \left\{ v_i | Z(v_i) = m, E(\Phi e_i) = n \right\},
  \end{equation}
and its second-level centroid $u_m^n$ as:
  \begin{equation}
  u_m^n = u_m + \Phi ^{-1} u^n,
  \end{equation}
where $\Phi ^{-1}$ is the inverse of the orthonormal matrix.
Similarly, we apply Eq. (\ref{eq:residual_ranking}) to generate the candidate set from the top-\textit{S} second-level clusters $\left\{ C_{m_r}^{n_s} \right\}$ from each $C_{m_r}$.

\subsubsection{Discussion}
\label{sec:codebook_size}

We have demonstrated the proposed probability-based ranking model can be integrated easily with the quantization methods described above.
Both RVQ and OPQ exploit the residual space where less quantization loss occurs.
Under the same quantization levels, OPQ has a more compact codebook size $O \big( (M+\beta)l \big)$ compared with RVQ, which takes $O \big( (M+N)l \big)$, for $\beta$ is usually smaller than \textit{N}.
The tradeoff is that the quantization loss is higher with OPQ than that using RVQ.

In our implementation of the two-level index structure, one first-level codebook and one second-level codebook are constructed, and the second-level codebook is shared by all the first-level clusters.
Alternatively, a second-level codebook can be constructed for each first-level cluster using locally-optimized quantization methods, such as hierarchical \textit{k}-means (HKM) \cite{nister2006scalable} and LOPQ \cite{kalantidis2014locally}.
These methods can lead to lower quantization losses (compared with RVQ and OPQ), but they must keep a large amount of second-level codebooks, e.g., $O(MNl)$ for HKM and $O(M \beta l)$ for LOPQ, which is not negligible in memory consumption.

\subsection{Hierarchical Learning}

The proposed ranking model can be applied hierarchically to learn the neighborhood relationships embedded in a hierarchical index space.
Again, we consider the two-level index structure of RVQ for illustration.
In Eq. (\ref{eq:residual_ranking}), we rank the second-level clusters according to their Euclidean distances from the query.
It is natural to consider replacing the distance-based ranking with the probability-based ranking for the second-level clusters.
However, it is impractical to generate the \textit{M} functions (of the \textit{M} first-level clusters) for the second-level probability prediction.
Instead, we propose only generating one function that can be shared for all of the first-level clusters.

We train a mapping function \textit{h} that maps a cluster-based feature $X_m$ to NN probabilities: $\left\{ p_m^1, p_m^2, ... , p_m^N \right\}$, where $p_m^n$ represents the NN probability of the second-level cluster $C_m^n$:
  \begin{equation} \label{eq:func_h}
  h(X_m) = \left\{ p_m^1, p_m^2, ... , p_m^N \right\}.
  \end{equation}
Let $X_m^{(t)}$ be the $C_m$-based feature of the \textit{t}th query $q^{(t)}$ in the training set.
$X_m^{(t)}$ comprises two different parts, which characterize the first-level and second-level clusters with respect to $q^{(t)}$.
The first part is the \textit{m}th first-level centroid $u_m^{(t)}, m \in \{1, 2, ... , M\}$, and the second part is the residual $e_m^{(t)} = q^{(t)} - u_m^{(t)}$.
$X_m^{(t)}$ is defined as the concatenation of the two parts $X_m^{(t)} = \left( u_m^{(t)}, e_m^{(t)} \right)$.
Therefore, $X_m^{(t)}$ generally characterizes the query representation in the second-level index space of the \textit{m}th first-level cluster $C_m$.
The target used in the training process is denoted as $Y_m^{(t)} = \left \{ y_m^{1(t)}, y_m^{2(t)}, ..., y_m^{N(t)} \right \}$, where $y_m^{n(t)}$ is the ground truth NN probability of the second-level cluster $C_m^n$:
  \begin{equation} \label{eq:target_y_m'}
  y_m^{n(t)} = \frac{\sum\limits_{k} \left\{ w_k^{(t)} \big | g_k^{(t)} \in C_m^n \right\}}{\sum\limits_{k} \left\{ w_k^{(t)} \big | g_k^{(t)} \in C_m^n \right\} + |C_m^n|}.
  \end{equation}
Then, the sum-to-one normalization is applied to $y_m^{n(t)}$.

It should be noted that only one mapping function \textit{h} is learned to predict the NN probabilities in the second-level index space.
Function \textit{h} is modeled by another fully-connected neural network and trained with the same procedure as function \textit{f}.
Suppose \textit{h} is characterized by the same numbers of hidden layers and units as \textit{f}.
The space complexity for functions \textit{f} and \textit{h} still requires $O(\tau\lambda^2 + \lambda M)$.
The time complexity through the two-level prediction takes $O \big( R(\tau\lambda^2 + \lambda M) \big)$, where the top-\textit{R} first-level clusters utilize \textit{h} for the second-level prediction with \textit{R} times in total.
%It appears that a large training data set $\left\{ \left( X_m'^{(t)}, Y_m'^{(t)} \right) \big | m = 1, 2, ... , M \right\}$ must be prepared for each $q^{(t)}, t \in \{ 1, 2, ... , T \}$.
%However, only a few data pairs $\left( X_m'^{(t)}, Y_m'^{(t)} \right)$ are useful for training because large numbers of the first-level clusters contain a small fraction of the NNs of the ground truth, and their target values defined in Eq. (\ref{eq:target_y_m'}) are usually close to zero, so these training data do not contribute greatly to training process.
%In practice, we set a criterion that if $\sum_{k} \big | y_m^{n(t)} \big |$ is less than a predefined threshold $\theta$, $\left( X_m'^{(t)}, Y_m'^{(t)} \right)$ will be excluded from the training data set.
%The compact training set makes the training process more efficient.

\section{Experimental Results}

In our experiments, we evaluated the proposed probability-based ranking and compared it to the distance-based ranking with various configurations and state-of-the-art methods.

\subsection{Datasets}

We examined based on two benchmark datasets comprising SIFT1B and DEEP1B.
SIFT1B is a visual feature set from the BIGANN Dataset \cite{jegou2011searching}, which has been used widely in many large-scale NN search studies.
SIFT1B contains one billion 128-dimensional SIFT descriptors and extra 10000 descriptors as the query set.
For each query, SIFT1B provides 1000 NNs with the smallest Euclidean distances from the query as the ground truth.
DEEP1B is another visual feature set \cite{babenko2016efficient}, which was produced by using the GoogLeNet architecture based on the ImageNet dataset.
The deep descriptors were extracted from the outputs of the last fully-connected layer, compressed using principal component analysis (PCA) to 96 dimensions, and normalized to unit length.
DEEP1B also contains one billion base descriptors and another 10000 as the query set.
Each query only provides one NN as the ground truth, so we extended it to 1000 NNs by running exhaustive search. \footnote{The 1000 NNs of the ground truth in DEEP1B and our source code project are available at \url{https://github.com/AmorntipPrayoonwong/Learning-to-Index-for-Nearest-Neighbor-Search}.}

\subsection{Implementation}

The proposed method uses the PCA-compressed data for indexing but the original data for reranking.
The basic idea is the “coarse-to-fine” approach.
Therefore, in the indexing stage we perform rough but efficient filtering with the compressed data, while in the reranking stage we execute ADC on the original data for detailed verification.
PCA, which is a standard tool for dimension reduction \cite{jegou2010aggregating}\cite{gong2013iterative}\cite{babenko2014neural}, was applied here to generate the compressed data for efficient indexing.
The two benchmark datasets were transformed from the original space to a 32-dimensional PCA-compressed space as the reference datasets.
We found that the indexing performance is even improved when using the compressed data instead of the original data, as discussed later.
On the other hand, reranking by ADC should be performed in the original space to avoid the transformation error.
We implemented ADC as follows.
The original SIFT/DEEP descriptor was partitioned into 16 segments.
Each segment was quantized by a codebook of 256 centroids, which were produced using \textit{k}-means clustering.
Since each segment could be encoded by 8 bits, the SIFT/DEEP descriptor was compressed to a 16-byte PQ code.
%When a query was given, it was partitioned into segments of 16-dimensional vectors, denoted as $q^j$ for the \textit{j}th segment.
%We created the \textit{j}th distance table $B^j = \left\{ \left\| q^j - u_m^j \right\|_2^2, m = 1, 2, ... , 256 \right\}$, where $u_m^j$ is the \textit{m}th centroid of the \textit{j}th codebook.
%The asymmetric distance between \textit{q} and any data point $v_i$ was calculated as:
%  \begin{equation}
%  ADC(q, v_i) = \sum_{j} \left\| q^j - u_{m'}^j \right\|_2^2,
%  \end{equation}
%where $u_{m'}^j$ is the nearest centroid of $v_i^j$.
%The calculations were performed efficiently by table lookup based on all of the distance tables.

For the proposed ranking model, mapping functions \textit{f} and \textit{h} were characterized based on a three-layer fully-connected neural network, expressed as: I1 - H2 - H3 - O4.
Function \textit{f} learned the neighborhood relationships in the first-level index space.
In its network, I1 was the input layer that received the query-dependent feature, and the number of input units was equal to the dimensions of the query-dependent feature.
H2 and H3 were hidden layers, where each had 512 units.
RELU was used as the activation function.
O4 was the output layer for predicting the NN probabilities of the first-level clusters and the number of output units was equal to the number of the first-level clusters.
The activation function was softmax and the loss function was cross entropy.
In addition, function \textit{h} characterized the relationships in the second-level index space and its network was slightly different from \textit{f} in layers I1 and O4.
I1 received the first-level centroid and residual of the query, and O4 predicted the NN probabilities of the second-level clusters.
Layers H2 and H3 had the same setup.

To train the neural network, for each benchmark dataset, we generated a training set containing 20000 data points, which were randomly selected with stratified sampling according to the data distribution over clusters.
Each training data point was associated with its 1000 NNs of ground truth found in the reference dataset.
%Generating correct NN lists by exhaustive search is time-consuming, so we applied ADC to obtain an approximate NN list for each training data point.
We set the batch size to 1000 and run 300 epochs for training.
The 10000 queries provided by the benchmark dataset were used for test.

The NN search methods were implemented in C++ and BLAS routines in the single-thread mode.
The neural networks were trained using Python and the Keras library.
The programs were run on a system that operated Windows 10 with an Intel Core i7 3.4 GHz CPU and 64 GB RAM.

\subsection{Results and Discussion}

First, we assess the quality of the candidate set retrieved from the first-level and second-level clusters.
The quantization methods RVQ and OPQ described above were investigated, where they were deployed in the two-level hierarchical index structures with various configurations, as listed in Table \ref{tb:index_structure}.
We denote the configuration as $M \times N$, where \textit{M} and \textit{N} are the numbers of the first-level clusters and second-level clusters, respectively.
For example, in the DEEP1B column, "$4096 \times 256$" indicates that the index structure used in DEEP1B had 4096 first-level clusters, each with 256 second-level clusters.
%The \textit{M} first-level clusters were generated in the 32-dimensional PCA-compressed space by \textit{k}-means clustering.
%The residual space characterized by the residual vectors of the reference data points was partitioned using \textit{N} codewords according to the quantization method. 
%In RVQ, \textit{k}-means clustering is applied based on the residual space to generate \textit{N} codewords directly.
%In OPQ, the residual space is divided into two subspaces, where each generates $\sqrt[]{N}$ codewords by non-parametric optimization \cite{ge2013optimized} and the overall quantization levels are still \textit{N} codewords in the residual space.
%IMI was applied to index the Cartesian product of the two subspaces \cite{babenko2012inverted}.

\begin{table}[ht]
\centering
\caption{Configurations of the index structures in the reference datasets}
\label{tb:index_structure}
\begin{tabular}{ |c|c| }
  \hline
  SIFT1B  &  DEEP1B \\
  \hline
  $256  \times 256$   &  $256  \times 256$ \\
  $1024 \times 1024$  &  $1024 \times 256$ \\
  $4096 \times 4096$  &  $4096 \times 256$ \\
  \hline
\end{tabular}
\end{table}

The accuracy was evaluated based on top-\textit{k} recall, which measures the quality of the retrieved set \textbf{A}.
Let $G_k = \{g_1, ..., g_k \}$ be the first \textit{k} NNs of ground truth for a query.
Top-\textit{k} recall is defined by:
  \begin{equation}
  \textit{top-k-recall}@\textbf{A} = \frac{| G_k \bigcap \mathbf{A} |}{| G_k |}.
  \end{equation}
In practice, we are often interested in the top-\textit{k} NNs ($k > 1$) rather than only the first NN ($k = 1$). However, as qualitative conclusions for $k = 1$ remain valid for $k > 1$, we apply top-1 recall in most of the experiments.

Note that the weights $\{ w_1, ..., w_K \}$ associated with $G_K$ are hyperparameters to be set according to the evaluation metric.
To obtain good top-1 recall, we will give a large weight to $w_1$.
If we are interested in top-\textit{k} recall, we can set higher values for $w_1, ..., w_k$.
In the following experiments, we set $w_1$ to 100 and other weights to 1 when top-1 recall is evaluated; for top-\textit{k} recall, we set $w_1, ..., w_k$ to 10 and other weights to 1.

\subsubsection{First-Level Retrieval}

In this experiment, we observe the effects on: (1) the first-level clusters ranked according to their NN probabilities instead of their distances from the query, and (2) the three query-dependent features used to derive the NN probabilities.
%Thus, we slightly modified Algorithm 1 as follows.
%Given the query \textit{q}, we ran Algorithm 1 from step 1 to step 3 to select the top-\textit{R} first-level clusters $\left\{ C_{m_r} | r = 1, 2, ..., R \right\}$.
%Note that step 4 was not applied here.
%Instead, we determined the number of the second-level clusters for retrieval in advance, which we denoted as $\overline{S}$, and this was fixed for every $C_{m_r}$.
%We set $\overline{S}$ = 5, 20, and 40 for different \textit{N} = 256, 1024, and 4096, respectively, under various $M \times N$ index structures.
%The reference data indexed in $\left\{ C_{m_r}^{n_s} | r = 1, 2, ..., R, s = 1, 2, ..., S \right\}$ were compiled as the candidate set \textbf{A} to check the top-1 recall for this retrieval.
To evaluate the performance, we considered the following approaches for comparison.

\begin{itemize}
  \item \emph{Distance-based ranking by query-centroid distances} (abbreviated as \textbf{Dist\_QCD}). This approach applies the conventional distance-based ranking to rank the first-level clusters according to their Euclidean distances from the query.
  \item \emph{Probability-based ranking by raw query data} (\textbf{Prob\_RAW}). This approach applies the proposed probability-based ranking to rank the first-level clusters according to their probabilities derived from the raw query data.
  \item \emph{Probability-based ranking by query-centroid similarities} (\textbf{Prob\_QCS}). This approach is similar to \textbf{Prob\_RAW} but the probabilities are derived from the similarities between the query and cluster centroids.
  \item \emph{Probability-based ranking by raw query data and query-centroid similarities} (\textbf{Prob\_RAW+QCS}). This approach is similar to \textbf{Prob\_RAW} but the probabilities are derived from the concatenation of the raw query data and query-centroid similarities.
  \item \emph{Ideal ranking} (\textbf{Ideal}). This approach foresees the NN ground truth distribution among clusters. It always select the best first-level clusters that contain the maximum of the NN ground truth.
\end{itemize}

\begin{figure*}[!htp]
\begin{center}
\includegraphics[width=1.0\textwidth]{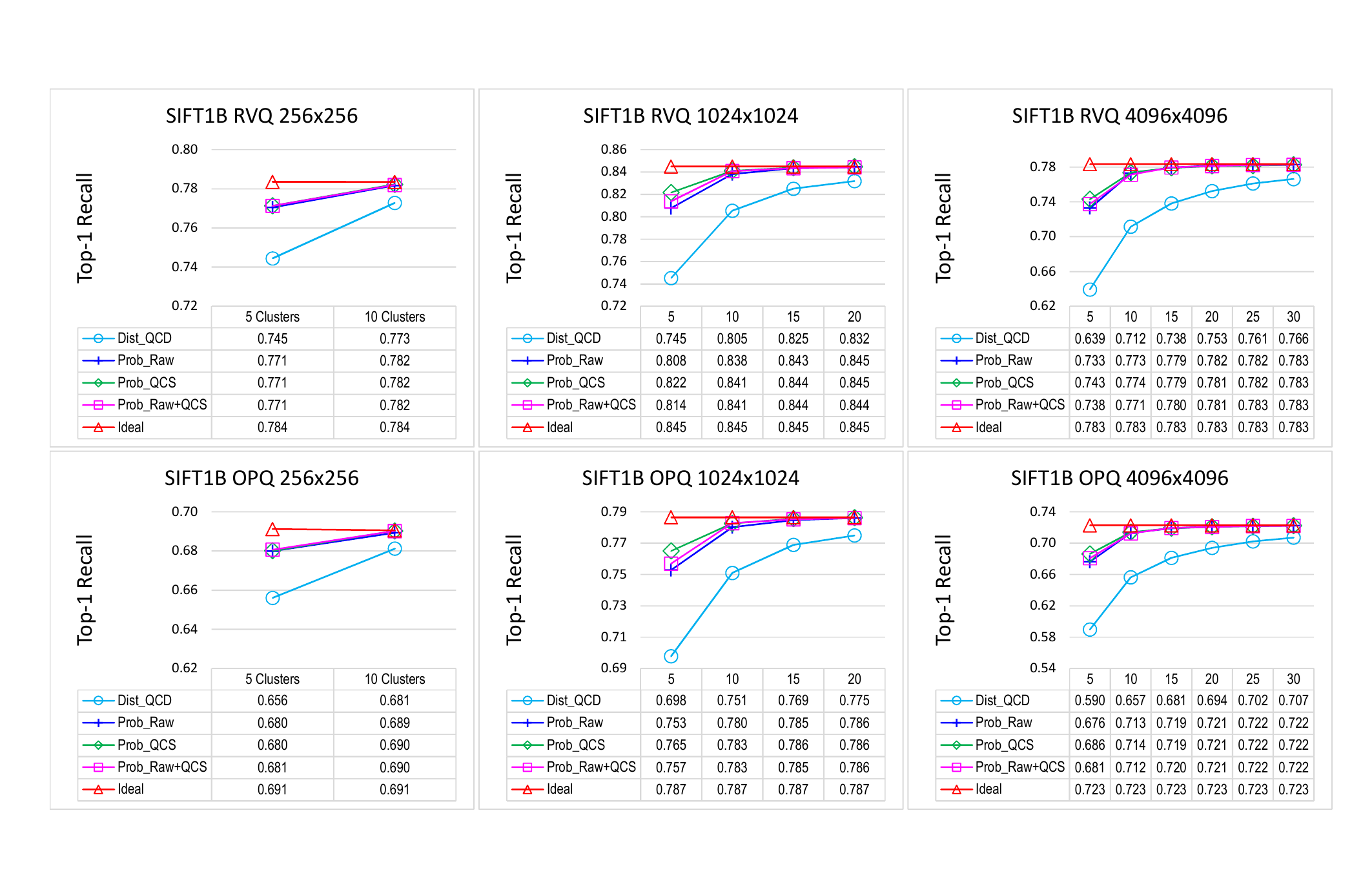}
\caption{Top-1 recall of the first-level retrieval in SIFT1B.}
\label{fig_1st_recall_sift}
\end{center}
\end{figure*}

\begin{figure*}[!htp]
\begin{center}
\includegraphics[width=1.0\textwidth]{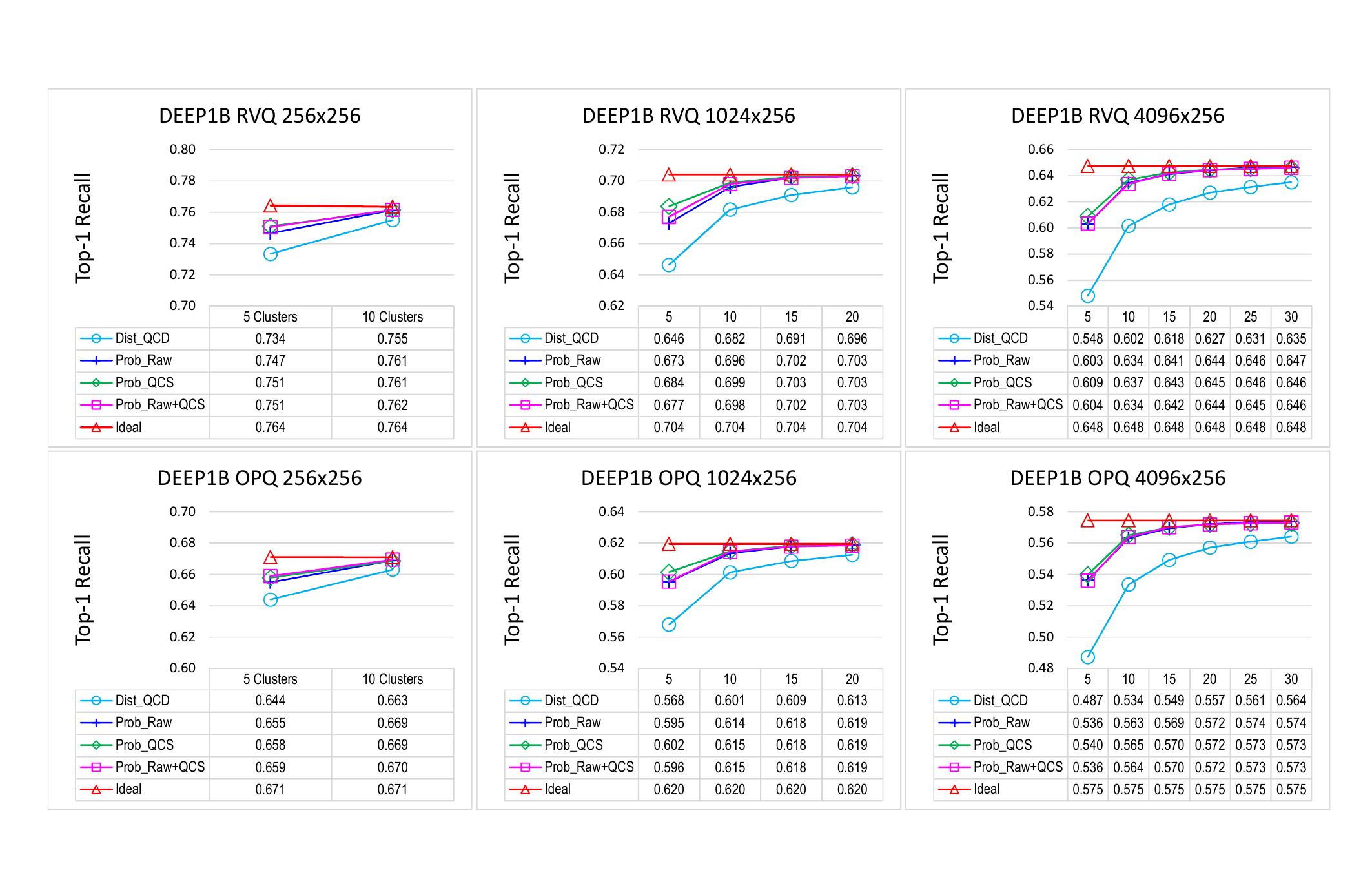}
\caption{Top-1 recall of the first-level retrieval in DEEP1B.}
\label{fig_1st_recall_deep}
\end{center}
\end{figure*}

%Tables \ref{tb:R@1K_SIFT1B} and \ref{tb:R@1K_DEEP1B} list the top-1 recall rates when retrieving the top-\textit{R} first-level clusters for SIFT1B and DEEP1B, respectively.
Figures \ref{fig_1st_recall_sift} and \ref{fig_1st_recall_deep} show top-1 recall when retrieving the top-\textit{R} first-level clusters for SIFT1B and DEEP1B, respectively, where \textit{R} = 5, 10, 15, 20, 25, and 30.
The \textbf{Ideal} approach obtained the highest accuracy and served as the upper bound.
The \textbf{Prob\_RAW}, \textbf{Prob\_QCS}, and \textbf{Prob\_RAW+QCS} approaches all obtained higher accuracy than the \textbf{Dist\_QCD} approach, thereby demonstrating that the proposed probability-based ranking generally performed better than the conventional distance-based ranking.
These results support our claim that the neighborhood relationships embedded in the index space can be learned by a nonlinear mapping function.
The function predicts the NN probabilities, which can be used in a more effective cluster ranking process than the Euclidean distances.

Among these probability-based approaches, \textbf{Prob\_QCS} obtained the best overall accuracy.
%It employs a higher-dimensional feature vector than \textbf{Prob\_RAW}.
%Intuitively, a neural network with more parameters can learn its non-linear function even better.
However, \textbf{Prob\_RAW+QCS} based on the concatenation of the two query-dependent features did not improve the performance further.
We consider that the query-centroid similarity feature provided sufficient information to learn a sufficiently good representation for NN search.
Concatenating the raw query data could introduce some noise or redundant information, and thus it had a detrimental influence on the search accuracy.

\subsubsection{Second-Level Retrieval}

In the next experiment we focus on the effects of quantity estimation and hierarchical learning for the second-level clusters.
We selected \textbf{Prob\_RAW} as the representative to demonstrate the ability of the proposed method.
\textbf{Prob\_RAW} did not have the best performance among the probability-based approaches, but it had the most compact neural network and this facilitated model training and inference.

%For each query, we retrieved the total \textit{T} second-level clusters from the top-\textit{R} first-level clusters.
%\textit{T} was determined by $T = R\cdot\overline{S}$, where $\overline{S}$ = 5, 20, and 40 for different \textit{N} = 256, 1024, and 4096, respectively.
%For example, with the DEEP1B $4096 \times 256$ dataset ($M = 4096, N = 256$), if we retrieved the top-10 first-level clusters ($R = 10$), then \textit{T} was determined by $T = 10 \times 5 = 50$, i.e., 50 second-level clusters were retrieved from the top-10 first-level clusters.
The selection for the second-level clusters was the main issue to be addressed here.
Thus, the following approaches with different selection schemes were implemented.

\begin{itemize}
  \item \emph{Distance-based ranking by query-centroid distances} (\textbf{Dist\_QCD}).
  \item \emph{Distance-based ranking by short-list extraction} (\textbf{Dist\_SLE}). This approach implements the short-list extraction method \cite{babenko2016efficient} based on distance-based ranking.
  \item \emph{Probability-based ranking by raw query data} (\textbf{Prob\_RAW}).
  \item \emph{Probability-based ranking by raw query data with sum normalization} (\textbf{Prob\_RAW+$\gamma_{sum}$}). This approach is based on \textbf{Prob\_RAW} but calculates \textit{S} by Eq. (\ref{eq:num_sec_cluster}) to determine the number of the second-level clusters to be retrieved. \textit{S} is rescaled by the sum normalization function given in Eq. (\ref{eq:sum_norm}) and  the second-level clusters are ranked by Eq. (\ref{eq:residual_ranking}).
  \item \emph{Probability-based ranking by raw query data with standard normalization} (\textbf{Prob\_RAW+$\gamma_{std}$}). This approach is similar to \textbf{Prob\_RAW+$\gamma_{sum}$} but \textit{S} is rescaled by the standard normalization function in Eq. (\ref{eq:std_norm}).
  \item \emph{Probability-based ranking by raw query data with exponent normalization} (\textbf{Prob\_RAW+$\gamma_{exp}$}). This approach is similar to \textbf{Prob\_RAW+$\gamma_{sum}$} but \textit{S} is rescaled by the exponent normalization function in Eq. (\ref{eq:exp_norm}).
  \item \emph{Hierarchical probability-based ranking by raw query data with standard normalization} (\textbf{HProb\_RAW+$\gamma_{sum}$}). This approach is similar to \textbf{Prob\_RAW+$\gamma_{sum}$} except hierarchical learning is applied. We rank the second-level clusters using Eq. (\ref{eq:func_h}) instead of Eq. (\ref{eq:residual_ranking}).
  \item \emph{Hierarchical probability-based ranking by raw query data with standard normalization} (\textbf{HProb\_RAW+$\gamma_{std}$}). This approach is similar to \textbf{Prob\_RAW+$\gamma_{std}$} except hierarchical learning is applied.
  \item \emph{Hierarchical probability-based ranking by raw query data with exponent normalization} (\textbf{HProb\_RAW+$\gamma_{exp}$}). This approach is similar to \textbf{Prob\_RAW+$\gamma_{exp}$} except hierarchical learning is applied.
\end{itemize}

%Suppose that the top-\textit{R} first-level clusters $\left\{ C_{m_r} | r = 1, 2, ..., R \right\}$ are given.
%\textbf{Dist\_QCD} and \textbf{Prob\_QCS} both select the top-$\overline{S}$ second-level clusters from each $C_{m_r}$, where $\overline{S}$ is the same for every $C_{m_r}$, and $T = R \overline{S}$ second-level clusters are gathered.
%\textbf{Dist\_SLE} aggregates all of the second-level clusters indexed in each $C_{m_r}$ in a pool and $RN$ second-level clusters are collected.
%From the pool, the top-\textit{T} second-level clusters are then selected according to their query-centroid distances.
%\textbf{Prob\_RAW+$\gamma_{sum}$}, \textbf{Prob\_RAW+$\gamma_{std}$}, and \textbf{Prob\_RAW+$\gamma_{exp}$} calculate \textit{S} dynamically for each $C_{m_r}$ and retrieve the top-\textit{S} second-level clusters from $C_{m_r}$.

\begin{figure*}[htp]
\begin{center}
\includegraphics[width=1.0\textwidth]{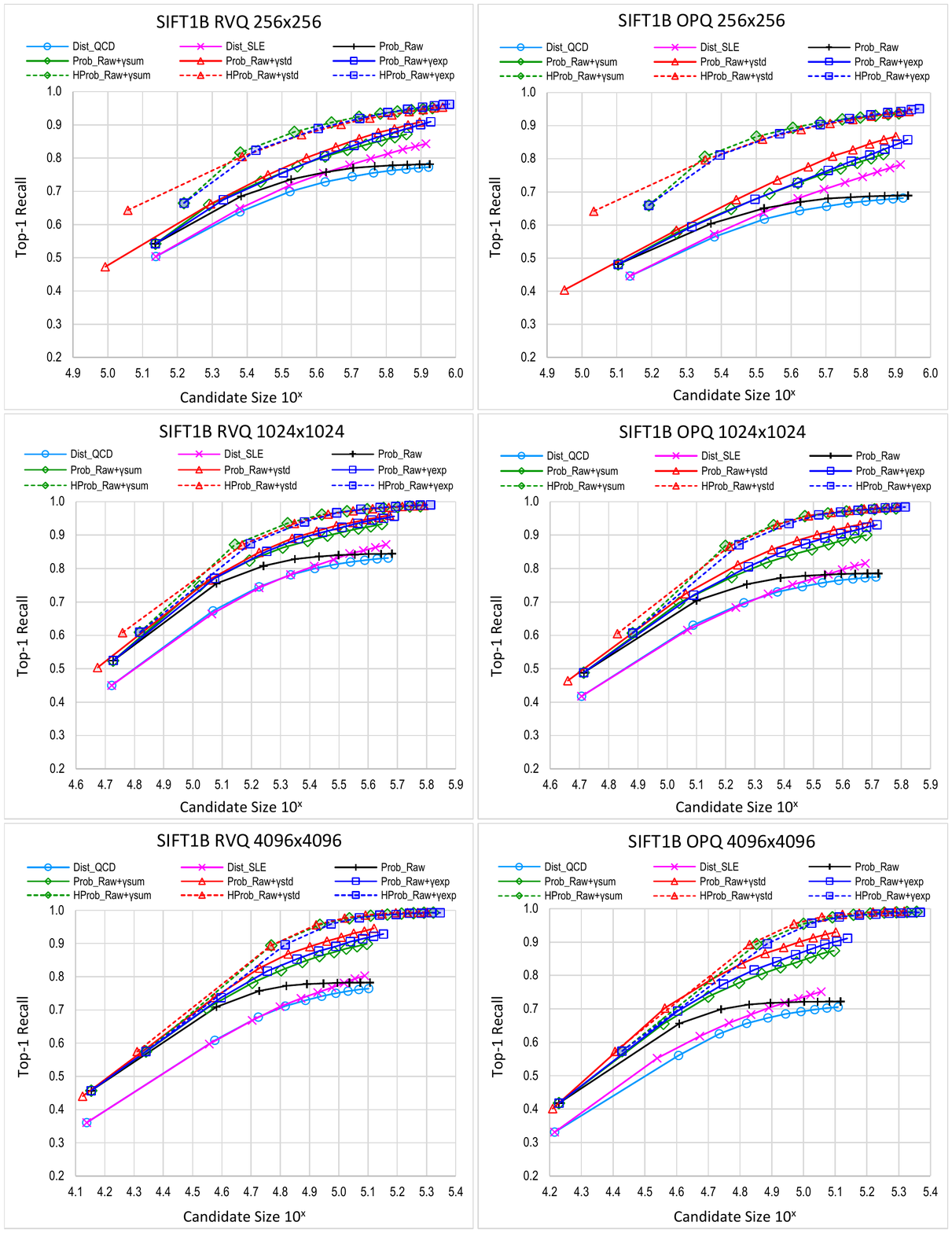}
\caption{Top-1 recall of the second-level retrieval in SIFT1B.}
\label{fig_2nd_recall_sift}
\end{center}
\end{figure*}

\begin{figure*}[htp]
\begin{center}
\includegraphics[width=1.0\textwidth]{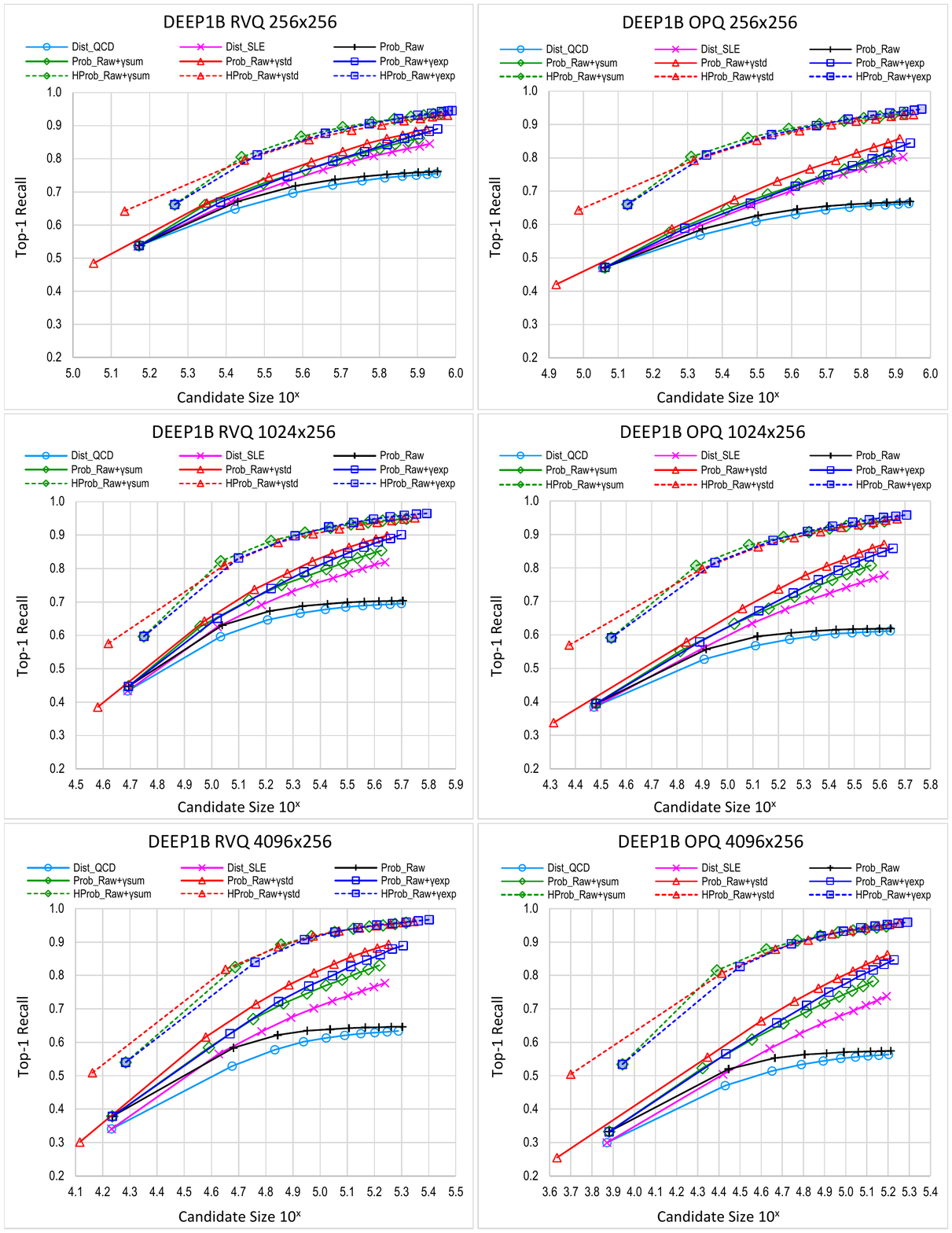}
\caption{Top-1 recall of the second-level retrieval in DEEP1B.}
\label{fig_2nd_recall_deep}
\end{center}
\end{figure*}

Figures \ref{fig_2nd_recall_sift} and \ref{fig_2nd_recall_deep} show top-1 recall along with the candidates retrieved from the second-level clusters for SIFT1B and DEEP1B, respectively.
The curves were generated with increasing numbers of the first-level clusters \textit{R}.
From the top-\textit{R} first-level clusters, these approaches retrieved the same number of second-level clusters but they were selected in different ways.
The result shows that, the proposed cluster quantity estimation effectively improved the search accuracy, thereby supporting our idea that the high-ranking cluster should retrieve a large number of candidates in proportion to its NN probability.
It makes the probability-based ranking outperforms the distance-based ranking.
Furthermore, the proposed hierarchical learning scheme greatly boosted the performance.
A surprising result was that the probability-based ranking still performed well at predicting the NN probabilities of subclusters when using only one neural network to characterize the neighborhood relationships in the residual (second-level) space.

The normalization functions $\gamma_{sum}$, $\gamma_{std}$, and $\gamma_{exp}$ all had beneficial effects when integrating with the ranking approach \textbf{Prob\_RAW}.
Function $\gamma_{std}$ obtained slightly better recall than the other two functions.
However, when applying hierarchical learning, they did not differ significantly and generally performed well in all cases.

\subsubsection{Scalability Analysis}
\label{sec:scalability_analysis}

We conduct three experiments to analyze the scalability of the proposed ranking model.
More precisely, we investigate the performance on a large size of the input/output layer of the neural network.

In the first experiment, we consider a large output layer size, which means a large number of clusters to be predicted.
We prepared the following first-level codebooks with different sizes \textit{M} = 256, 1024, 4096, and 16384 to produce the SIFT1B RVQ index structures for the first-level cluster retrieval.
The number of the second-level clusters \textit{N} was fixed to 256.
Table \ref{tb:codebook_size} lists the result by using the \textbf{Prob\_RAW} approach when retrieving the top-\textit{R} first-level clusters, \textit{R} = 1, 3, 5, 10, and 15.
The runtime overhead mainly comes from probability prediction and partial sorting for \textit{M} clusters, which takes $O(\tau\lambda^2 + \lambda M)$ and $O(M)$, respectively (see Section 3.3).
Hence, the runtime grows linearly with \textit{M}.

The second experiment is about the input dimensionality.
We made a comparison for indexing the PCA-compressed and original data in SIFT1B RVQ $4096\times4096$ and DEEP1B RVQ $4096\times4096$ datasets.
For the ranking model, we only changed the input layer size of the neural network to fit the compressed/original data, and kept other layer sizes unchanged.
Top-1 recall and runtime were evaluated by using the \textbf{HProb\_RAW+$\gamma_{std}$} approach to perform the second-level retrieval within one hundred thousand candidates.
As shown in Table \ref{tb:feature_size}, the proposed ranking model is scalable for high-dimensional input data.
More important, the original data can be compressed by PCA to enjoy a better indexing performance.

Table \ref{tb:top-k_recall} lists the third experimental result for top-\textit{k} recall, \textit{k} = 1, 10, 100, and 1000.
These results are yielded by using \textbf{HProb\_RAW+$\gamma_{std}$} in SIFT1B and DEEP1B RVQ $4096 \times 4096$.

\begin{table}[htp]
\centering
\caption{Top-1 recall and runtime (milliseconds) of the first-level retrieval for different first-level codebooks in SIFT1B RVQ}
\label{tb:codebook_size}
\begin{tabular}{ |c|c|c|c|c|c|c| }
  \hline
    &  1 clu.  &  3 clu.  &  5 clu.  &  10 clu.  &  15 clu.  &  time  \\
  \hline
  $256 \times 256$  &  0.612  &  0.637  &  0.637  &  0.637  &  0.637  &  0.06  \\
  $1024 \times 256$  &  0.583  &  0.708  &  0.721  &  0.722  &  0.722  &  0.09  \\
  $4096 \times 256$  &  0.644  &  0.783  &  0.799  &  0.804  &  0.804  &  0.20  \\
  $16384 \times 256$  &  0.717  &  0.851  &  0.864  &  0.870  &  0.871  &  0.57  \\
  \hline
\end{tabular}
\end{table}

\begin{table}[htp]
\centering
\caption{Top-1 recall and runtime (milliseconds) of the second-level retrieval for indexing PCA-compressed and original data in SIFT1B and DEEP1B RVQ $4096 \times 4096$}
\label{tb:feature_size}
\begin{tabular}{ |c|c|c|c|c|c|c| }
  \hline
    &  1 clu.  &  3 clu.  &  5 clu.  &  10 clu.  &  15 clu.  &  time  \\
  \hline
  SIFT1B 32d  &  0.803  &  0.963  &  0.984  &  0.995  &  0.996  &  1.52  \\
  SIFT1B 128d  &  0.779  &  0.956  &  0.978  &  0.990  &  0.991  &  1.58  \\
  DEEP1B 32d  &  0.756  &  0.946  &  0.975  &  0.991  &  0.993  &  1.52  \\
  DEEP1B 96d  &  0.753  &  0.943  &  0.972  &  0.988  &  0.992  &  1.56  \\
  \hline
\end{tabular}
\end{table}

\begin{table}[htp]
\centering
\caption{Top-\textit{k} recall (\textit{k} = 1, 10, 100, and 1000) in SIFT1B and DEEP1B RVQ $4096 \times 4096$}
\label{tb:top-k_recall}
\begin{tabular}{ |c|c|c|c|c| }
  \hline
    &  Top-1  &  Top-10  &  Top-100  &  Top-1000  \\
  \hline
  SIFT1B  &  0.996  &  0.646  &  0.598  &  0.506  \\
  DEEP1B  &  0.993  &  0.620  &  0.577  &  0.487  \\
  \hline
\end{tabular}
\end{table}

\subsubsection{Comparison to the State-of-the-Art}

Finally, we compare the proposed method with the results reported in the literature \cite{jegou2011searching}\cite{babenko2012inverted}\cite{kalantidis2014locally}\cite{babenko2016efficient}\cite{baranchuk2018revisiting}, as shown in Table \ref{tb:comparison}.
To make a fair comparison, we report the proposed method using the original data, instead of the PCA-compressed data, in the indexing stage.
The first column lists these methods and their index structure configurations.
The memory column measures the index structure size (in bytes per data point) and omits the 4-byte data identity and 16-byte PQ code.
Each method retrieved within one hundred thousand candidates per query for reranking.
The proposed method was evaluated using \textbf{HProb\_RAW+$\gamma_{std}$} in SIFT1B and DEEP1B RVQ $4096\times4096$.

\begin{table*}[!htp]
\centering
\caption{Top-1 recall@\textbf{A} (R@1, R@10, and R@100), runtime (milliseconds), and memory consumption (bytes) of the state-of-the-art methods}
\label{tb:comparison}
\begin{tabular}{ |c||c|c|c|c|c||c|c|c|c|c| }
  \hline
  \multirow{2}{*}{} & \multicolumn{5}{c||}{SIFT1B} & \multicolumn{5}{c|}{DEEP1B} \\
  \cline{2-11} 
                    &  R@1  &  R@10  &  R@100  &  time  &  mem  &  R@1  &  R@10  &  R@100  &  time  &  mem  \\
  \hline
  \hline
  IVFADC $(2^{20})$ \cite{jegou2011searching}  &  0.351  &  0.786  &  0.918  &  2.5  &  0.52  &  0.405  &  0.773  &  0.916  &  2.5  &  0.40  \\
  \hline                  
  IMI $(2^{14} \times 2^{14})$ \cite{babenko2012inverted}  &  0.360  &  0.792  &  0.901  &  5.0  &  1.34  &  0.397  &  0.766  &  0.909  &  8.5  &  1.34  \\
  \hline
  Multi-LOPQ $(2^{14})$ \cite{kalantidis2014locally}  &  0.454  &  0.862  &  0.908  &  19.0  &  3.22  &  0.410  &  0.790  &  -  &  20.0  &  2.68  \\
  \hline                  
  GNOIMI $(2^{14} \times 2^{14})$ \cite{babenko2016efficient}  &  -  &  -  &  -  &  -  &  -  &  0.450  &  0.810  &  -  &  20.0  &  3.75  \\
  \hline
  IVFOADC+GP $(2^{20})$ \cite{baranchuk2018revisiting}  &  0.405  &  0.851  &  0.957  &  3.5  &  2.00  &  0.452  &  0.832  &  0.947  &  3.3  &  1.87  \\
  \hline                  
  Proposed $(2^{12} \times 2^{12})$  &  0.457  &  0.862  &  0.977  &  4.7  &  0.16  &  0.475  &  0.785  &  0.938  &  4.5  &  0.16  \\
  \hline
\end{tabular}
\end{table*}

Our method demonstrated a comparable performance to the state-of-the-arts.
In particular, our method is memory-efficient for indexing.
It employs the following data structures: an index table, two codebooks, and two neural networks.
The two-level index table, which keeps $4096\times4096$ pointers to inverted lists, takes 128 MB.
The first-level and second-level codebooks, each of which stores 4096 128-dimensional real-valued vectors (for SIFT), require 8 MB.
The two neural networks used for hierarchical prediction occupy 17 MB.
In total we consume close to 160 MB memory to index one billion data points.

\begin{figure}[htp]
\begin{center}
\includegraphics[width=0.45\textwidth]{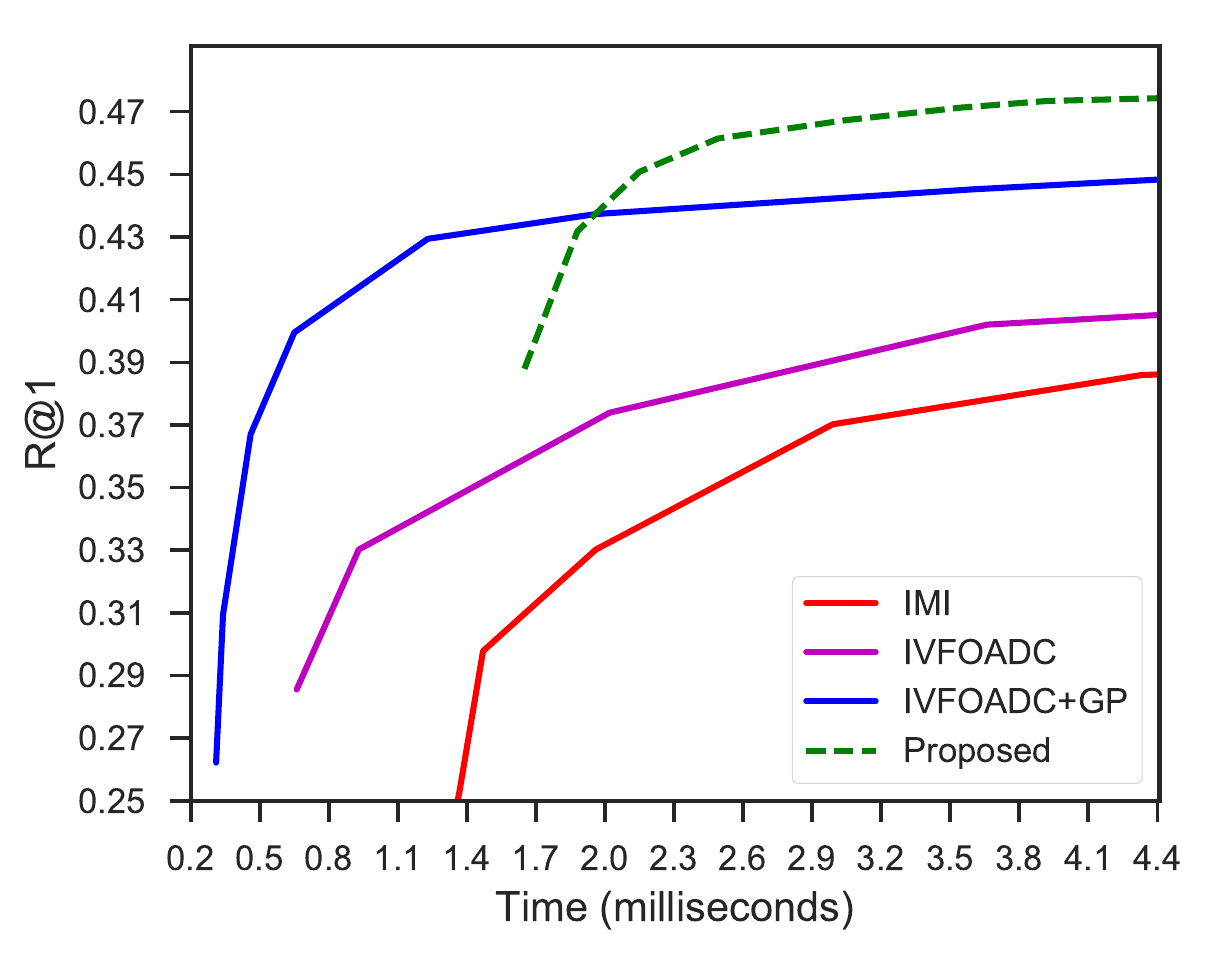}
\caption{R@1 vs. runtime in DEEP1B.}
\label{fig_r1_time}
\end{center}
\end{figure}

Note that in ADC, applying conventional PQ in DEEP1B is less effective than that in SIFT1B due to their different nature; special treatment can be adopted to improve the recall rates in DEEP1B \cite{babenko2016efficient}.
Figure \ref{fig_r1_time} plots R@1 vs. runtime in DEEP1B to compare the state-of-the-art methods.
The runtime for the proposed method started from 1.6 milliseconds, which included the computation overhead for starting to retrieve the first cluster.
After 2 milliseconds the proposed method yielded higher recall rates than the other methods.

\section{Conclusion}

In this study, we proposed a novel ranking model for learning the neighborhood relationships embedded in the index space.
The proposed model ranks clusters based on their NN probabilities predicted by the learned neural networks.
It can replace the distance-based ranking model and may be integrated with other quantization/search methods to boost their retrieval performance.
Our experimental results demonstrated that the proposed model was effective at alleviating the information loss problem during NN search.

\ifCLASSOPTIONcompsoc
  \section*{Acknowledgments}
\else
  \section*{Acknowledgment}
\fi

The authors would like to thank Dmitry Baranchuk for the kindly help with figure reproducing and the reviewers for the thoughtful comments and suggestions.

\ifCLASSOPTIONcaptionsoff
  \newpage
\fi



%\bibliographystyle{IEEEtran}

%\begin{thebibliography}{1}

%H.~Kopka and P.~W. Daly, \emph{A Guide to {\LaTeX}}, 3rd~ed.\hskip 1em plus

%\end{thebibliography}

\bibliographystyle{IEEEtran}
\bibliography{sample-bibliography}


\begin{IEEEbiography}[{\includegraphics[width=1in,height=1.25in,clip,keepaspectratio]{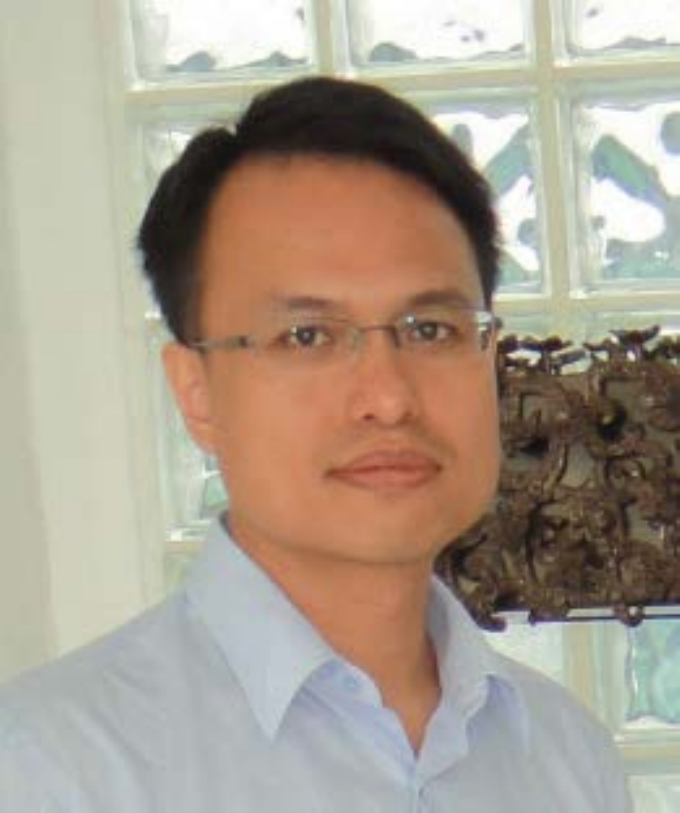}}]{Chih-Yi Chiu}
Chih-Yi Chiu received the B.S. degree in information management from National Taiwan University in 1997, and the M.S. degree in computer science from National Taiwan University in 1999, and the Ph.D. degree in computer science from National Tsing Hua University, Taiwan, in 2004. From 2005 to 2009, he was with Academia Sinica. In 2009, he joined National Chiayi University. His current research interests include multimedia information retrieval, data mining, and deep learning.
\end{IEEEbiography}

\begin{IEEEbiography}[{\includegraphics[width=1in,height=1.25in,clip,keepaspectratio]{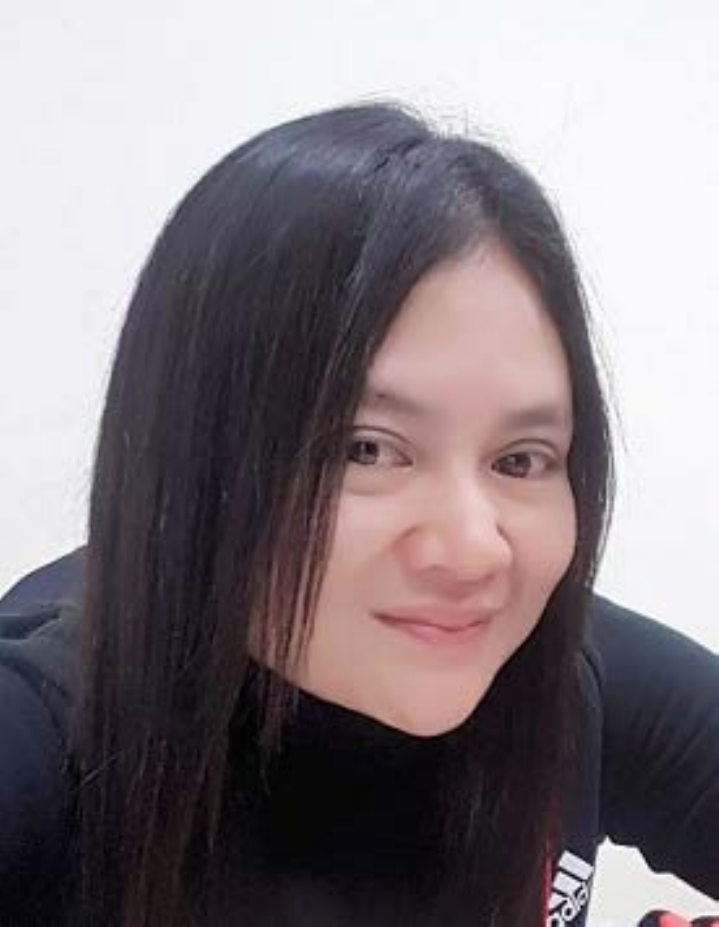}}]{Amorntip Prayoonwong}
Amorntip Prayoonwong received the B.S. degree in Business Computer from Yonok University, Thailand, in 1995, and the M.S. degree in Computer Science from Assumption University, Thailand, in 2004.
She is currently pursuing the Ph.D. degree in the Computer Science and Information Engineering, National Chiayi University.
Her current research interest includes index structures and nearest neighbor search.
\end{IEEEbiography}

\begin{IEEEbiography}[{\includegraphics[width=1in,height=1.25in,clip,keepaspectratio]{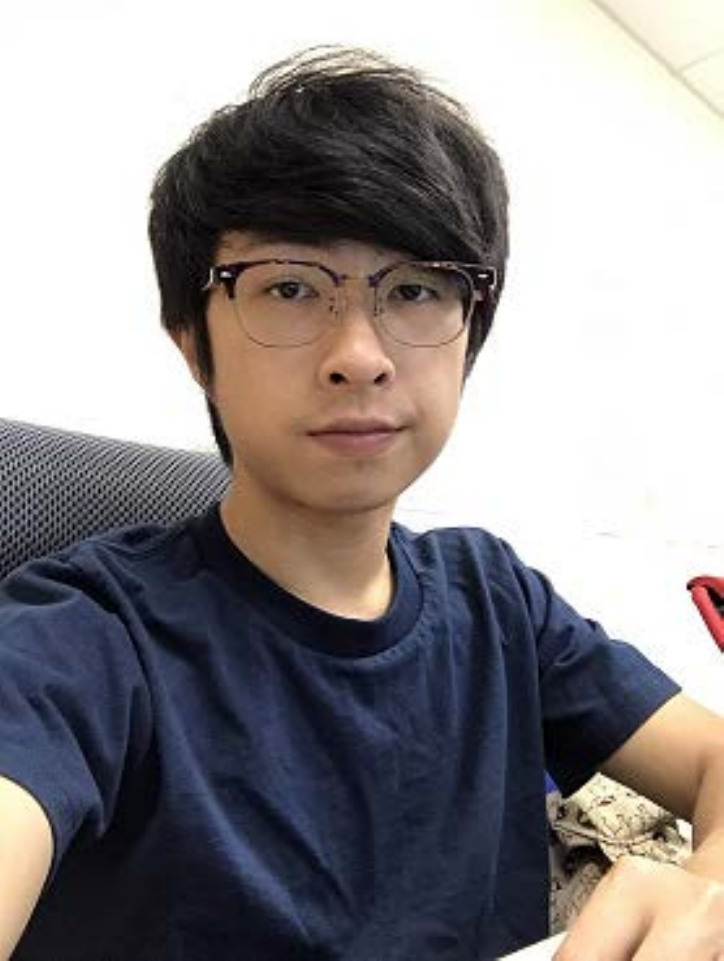}}]{Yin-Chih Liao}
Yin-Chih Liao received the B.S. degree in Computer Science and Information Engineering from National Chiayi University in 2017. His current research interests include multimedia information retrieval and deep learning.
\end{IEEEbiography}


%\vfill


\end{document}